\documentstyle[nato]{crckapb} 

\input epsf

\def\simlt{\stackrel{<}{{}_\sim}}
\def\simgt{\stackrel{>}{{}_\sim}}

\begin{opening}
\title{PARTICLE PHYSICS IN THE EARLY UNIVERSE}

\author{EDWARD W.\ KOLB}
\institute{Fermi National Accelerator Laboratory\\
          Batavia, Illinois 60510 \ \ \ USA}

\end{opening}

\begin{document}

\section{Introduction}

Perhaps the most striking illustration of the true unity of science is
the development of the interdisciplinary field of ``particle
cosmology.''  Particle physics examines nature on the smallest scales,
while cosmology studies the universe on the largest scales.  Although
the two fields are separated by the scales of the objects they study,
they are unified because it is impossible to understand the origin and
evolution of large-scale structures in the universe without
understanding the ``initial conditions'' that led to the structures.
The initial data was set in the very early universe when the
fundamental particles and forces acted to produce the perturbations in
the cosmic density field.  A complete understanding of the present
structure of the universe will also be impossible without accounting
for the dark component in the density field.  The most likely
possibility is that this ubiquitous dark component is an elementary
particle relic from the early universe.

The study of the structure of the present universe may reveal insights
into events which occurred in the early universe, and hence, into the
nature of the fundamental forces and particles at an energy scale far
beyond the reach of terrestrial accelerators.  Perhaps the early
universe was the ultimate particle accelerator, and will provide the
first glimpse of physics at the scale of Grand Unified Theories
(GUTs), or even the Planck scale.

As a cosmologist I am interested in events that happened a long time
ago.  But in studying the past, I believe it is best to take the
approach of a historian rather than an antiquarian.  Now an
antiquarian and a historian are both interested in things from the
past.  But an antiquarian is interested in old things just because
they are old.  To an antiquarian, there is no difference between a
laundry list from June 1215 and the Magna Carta: they are both equally
old.  A historian, on the other hand, is interested in the past
because it shapes the present.  The job of a historian is to sort
through events of the past and see which are important and which are
not.  I am not interested in the early universe just because it
happened a long time ago, or it was really hot, or it was a bang (a
really, really big one).  The real reason I study the early universe
is that events which occurred in the early universe left an imprint
upon the present universe.

In these lectures I will concentrate on two events which occurred in
the early universe.  The first is the generation of perturbation in
the density field during an early period of rapid expansion known as
cosmic inflation.  The second is the genesis of dark matter.  The
record of these events is written in the arrangement of galaxies,
galaxy clusters, and imperfection in the isotropy of the cosmic
microwave background radiation.  If we really understood particle
physics, we could predict the nature of those patterns.  If we really
knew how to read the story in the structures, we would learn something
about particle physics.  The story is there on the sky, patiently
waiting for our wits to become sharp enough to read it.

In these lectures I will discuss the early universe.  So the first
thing we must do is to follow the procedure outlined by William
Shakespeare \cite{bill}:
\begin{quote}
Now entertain conjecture of a time\\
When creeping murmur and the poring dark\\
Fills the wide vessel of the universe.
\end{quote}

\section{The Density Field of the Universe}

\def\vecx{{\vec{x}}}
\def\veck{{\vec{k}}}
\def\avg#1{{\langle #1 \rangle}}

The universe is not exactly homogeneous and isotropic, but it is a
sufficiently accurate description of the universe on large scales that
it is useful to consider homogeneity and isotropy as a first
approximation, and discuss departures from this idealized smooth
universe.

Let us begin by considering the density field, $\rho(\vecx)$.  If the
average density of the universe is denoted as $\avg\rho$, then we can
define a dimensionless density contrast $\delta(\vecx)$ as
\begin{equation}
\delta(\vecx) = \frac{\rho(\vecx)-\avg\rho } {\avg\rho} \ .
\end{equation}
Of course we cannot predict $\delta(\vecx)$, but we can hope to
predict the statistical properties of $\delta(\vecx)$.  The correct
arena to discuss the statistical properties of the density field is in
Fourier space, where one decomposes the density contrast into its
various Fourier modes $\delta_\veck$:
\begin{equation}
\delta(\vecx) = V \int d^3\!k \, \delta_\veck \, e^{-i\veck \cdot \vecx},
\end{equation}
where $V$ is some irrelevant normalization volume.  After a little
Fourier manipulation and some mild assumptions about the density
field, it is easy to show that the two-point autocorrelation function
of the density field can be expressed solely in terms of $\left|
\delta_\veck \right|^2$:
\begin{equation}
\avg{\delta(\vecx)\delta(\vecx)} = 
A\int_0^\infty \frac{dk}{k} \, k^3 \, \left| \delta_\veck \right|^2\ ,
\end{equation}
where $A$ is yet another irrelevant constant.

So long as the fluctuations are Gaussian, all statistical information
is contained in a quantity known as the {\it power spectrum}, which
can be defined as either
\begin{eqnarray}
\Delta^2(k ) & = & 
           k^3 \left| \delta_\veck \right|^2 \ , \qquad {\rm or} \nonumber \\
P(k) & = & \left| \delta_\veck \right|^2 \ .
\end{eqnarray}
The first choice is much more physical, as it represents the power per
logarithmic decade in the fluctuations.  Although the first choice
makes much more sense, the second choice is what is usually used.  It
turns out that graphs of $P(k)$ have a nicer form (but less physical
content) than corresponding graphs of $\Delta^2(k)$.  Since the
widespread availability of color graphics, presentation seems to be
everything, and information content of secondary concern.

\subsection{The power spectrum from large-scale structure}

The power spectrum  is related to  the {\it rms} fluctuations in the density
on  scale  $R=2\pi/k$.  The exact  relationship  depends upon sampling
procedure,  window functions, {\it etc.}   But for a  simple intuitive feel,
imagine we have mass points spread throughout some sample volume.  Now
place  a sphere of  radius $R$ in the volume   and count the number of
points within the sphere.  Then  repeat as often as  you have the time
or patience to do so.  There will be an  average number $\avg{N}$, and
an {\it rms} fluctuation  $\avg{  (\delta  N/\avg{N})^2}^{1/2}$.   The power
spectrum is related to that   {\it rms} fluctuation: $\Delta(k=2\pi  R^{-1})
\propto \avg{  (\delta N/\avg{N})^2}^{1/2}$.  Repeating  the procedure
for many values of $R$ will  give $\Delta$ as  a function of $R$---the
power spectrum.

Now how does one go about observing the mass within a sphere of radius
$R$?  Well, it is difficult to measure the mass.  It is easier to
count the number of galaxies.  So one assumes that the galaxy
distribution traces the mass distribution.  Although it seems
reasonable that regions of high density of galaxies correspond to
regions of high mass density, since most of the mass is dark, the
proportionality might not be exact.  Thus, we have to allow for a
possible {\it bias} in the power spectrum.  Other problems also arise.
The distance to an object is not measured directly; what is measured
is its redshift.  The redshift is determined by the distance to the
object, as well as its peculiar motion.  In regions of large
overdensity the peculiar motions may be large, resulting in what is
known as redshift distortions.  Another problem is nonlinear
evolution, which distorts the power spectrum in regions of large
overdensity.  Thus, if one wants to compare the observed power
spectrum with the linear power spectrum generated by early-universe
physics, it is necessary to make corrections for bias, redshift-space
distortions, and nonlinear evolution.

\begin{figure}[t]
\centering
\leavevmode\epsfxsize=350pt  \epsfbox{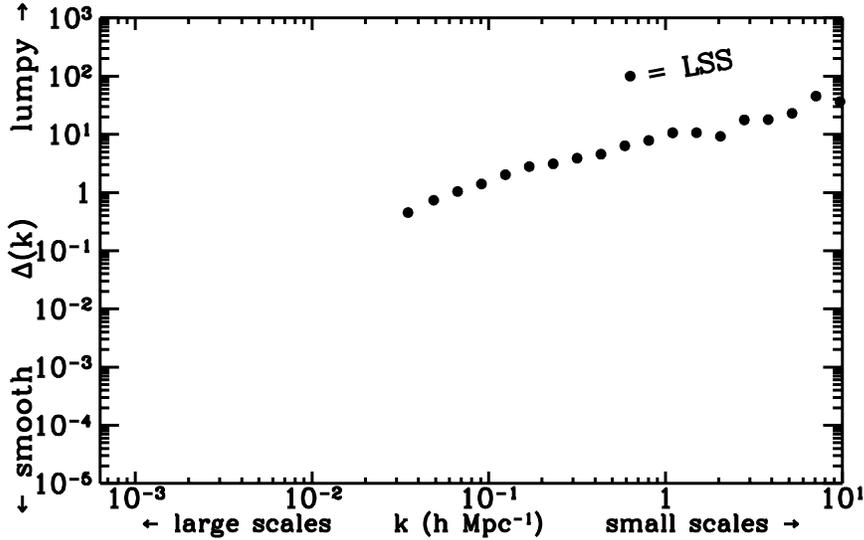}
\caption[fig1]{\label{p_k_lss} An example of a power spectrum deduced
from a large-scale structure (LSS) survey.  A megaparsec (Mpc) is
$3.26\times10^{24}$cm, and $h$ is the dimensionless Hubble constant,
$H_0= 100h$ km s$^{-1}$Mpc$^{-1}$.  A wavenumber $k$ is roughly
related to a length scale of $2\pi/k$. }
\end{figure}

Deducing the power spectrum from galaxy surveys is a tricky business.
Rather than go into the details, uncertainties, and all that, I will
just present a representative power spectrum in Fig.\ (\ref{p_k_lss}).
Note that $\Delta(k)$ decreases with increasing length scale
(decreasing wavenumber).  The universe is lumpy on small scales, but
becomes progressively smoother when examined on larger scales.

\newpage

\subsection{The power spectrum deduced from the cosmic background radiation}

The microwave background is isotropic to about one part in $10^3$.  If
one removes the anisotropy caused by our motion with respect to the
cosmic background radiation (CBR) rest frame, then it is isotropic to 
about 30 parts-per-million.
But as first discovered by the Cosmic Background Explorer (COBE),
there are intrinsic fluctuation in the temperature of the CBR.

\begin{figure}[t]
\centering
\leavevmode\epsfxsize=350pt  \epsfbox{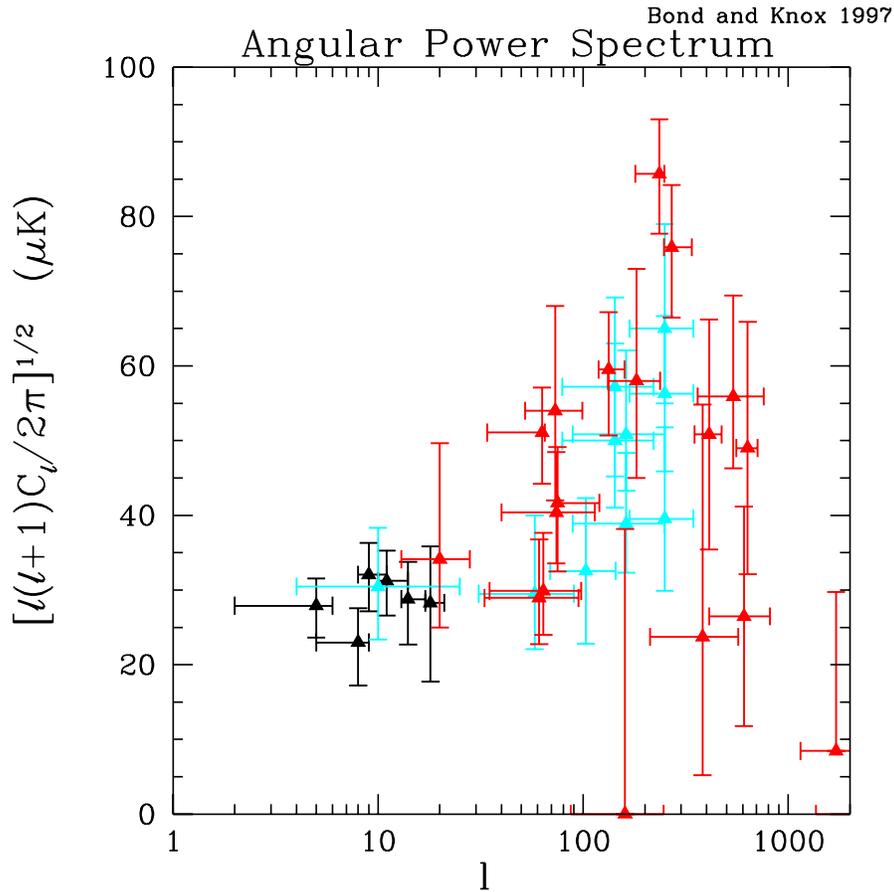}
\caption[fig2]{\label{cl} The angular power spectrum of CBR
fluctuations (courtesy of Dick Bond and Llyod Knox).}
\end{figure}

Just as perturbations in the density field were expanded in terms of
Fourier components, a similar expansion is useful for temperature
fluctuations.  Because the surface of observation about us
can be described in terms of spherical angles $\theta$ and $\phi$,
the correct expansion basis is spherical harmonics,
$Y_{lm}(\theta,\phi)$.  If the average temperature is $\avg{T}$, then
one can expand
\begin{equation}
\frac{\Delta T(\theta,\phi)}{\avg{T}} = 
\sum_{l,m} a_{lm} Y_{lm}(\theta,\phi) \ .
\end{equation}

Of course $\avg{a_{lm}}=0$, but with proper averaging, 
\begin{equation}
\avg{\left|a_{lm}\right|^2}  \equiv C_l  \neq 0 \ .
\end{equation}
$C_l$ as a function of $l$ is called the angular power spectrum.  In
the six years since the first measurement of CBR fluctuations by COBE,
a number of experiments have detected
fluctuations.  The present situation is illustrated in Fig.\ \ref{cl}.

Associated with a multipole number $l$ is a characteristic angle
$\theta$, and a length scale we can define as the distance subtended
by $\theta$ on the surface of last scattering.  Since the distance to
the last scattering surface of the microwave background is so large,
the temperature fluctuations represent the largest structures ever
seen in the universe.

Contributing to the temperature anisotropies are fluctuations in the
gravitational potential on the surface of last scattering.  Photons
escaping from regions of high density will suffer a larger than
average gravitational redshift, hence will appear to originate from a
cold region.  In similar fashion, photons coming to us from a
low-density region will appear hot.  In this manner, temperature
fluctuations can probe the density field on the surface of last
scattering and provide information about the power spectrum on scales
much larger than can be probed by conventional large-scale structure
observations.

\begin{figure}[p]
\centering
\leavevmode\epsfxsize=350pt  \epsfbox{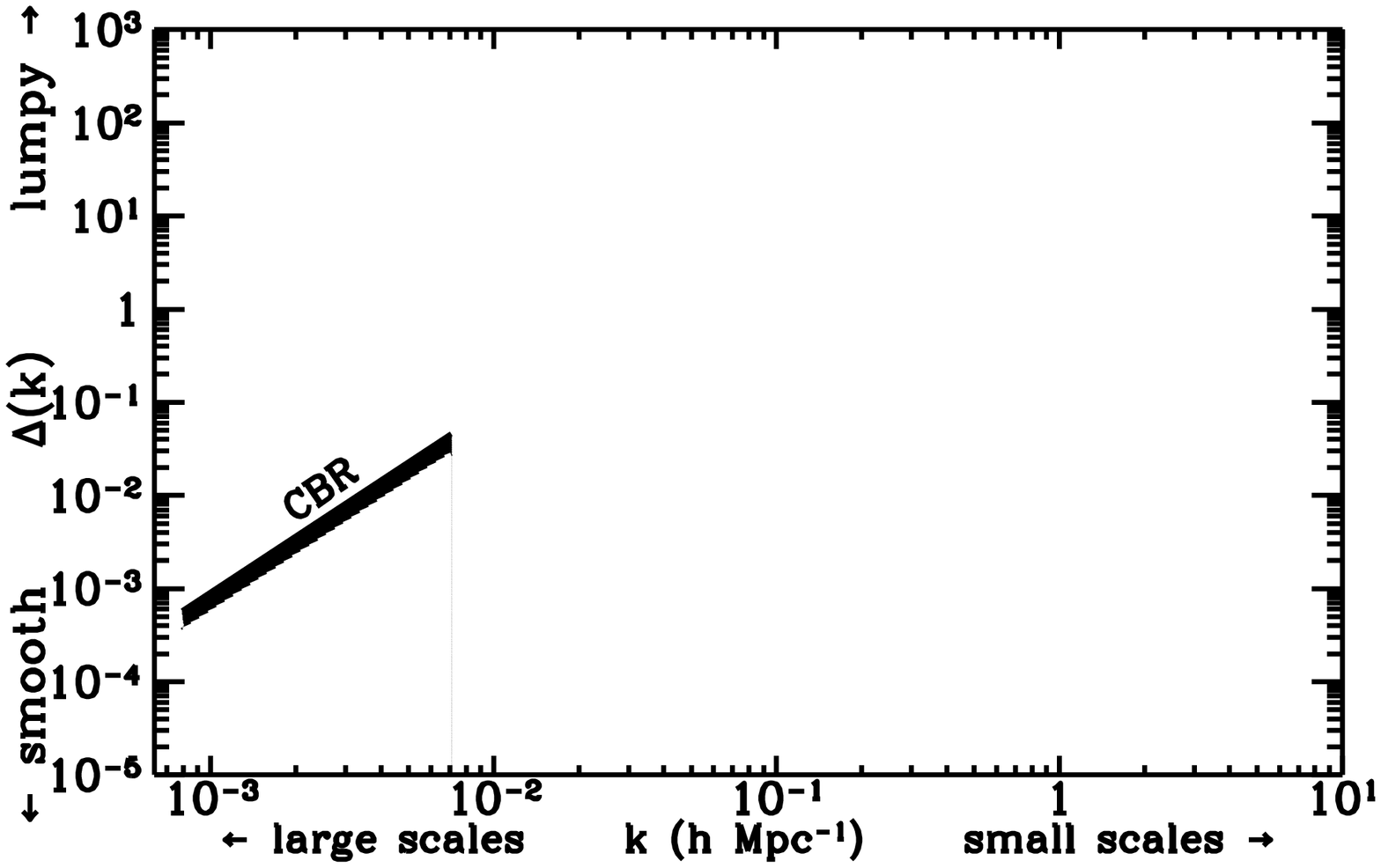}
\caption[fig3]{\label{p_k_cbr} The power spectrum deduced by
measurements of large angular scale CBR temperature fluctuations.}
\leavevmode\epsfxsize=350pt \epsfbox{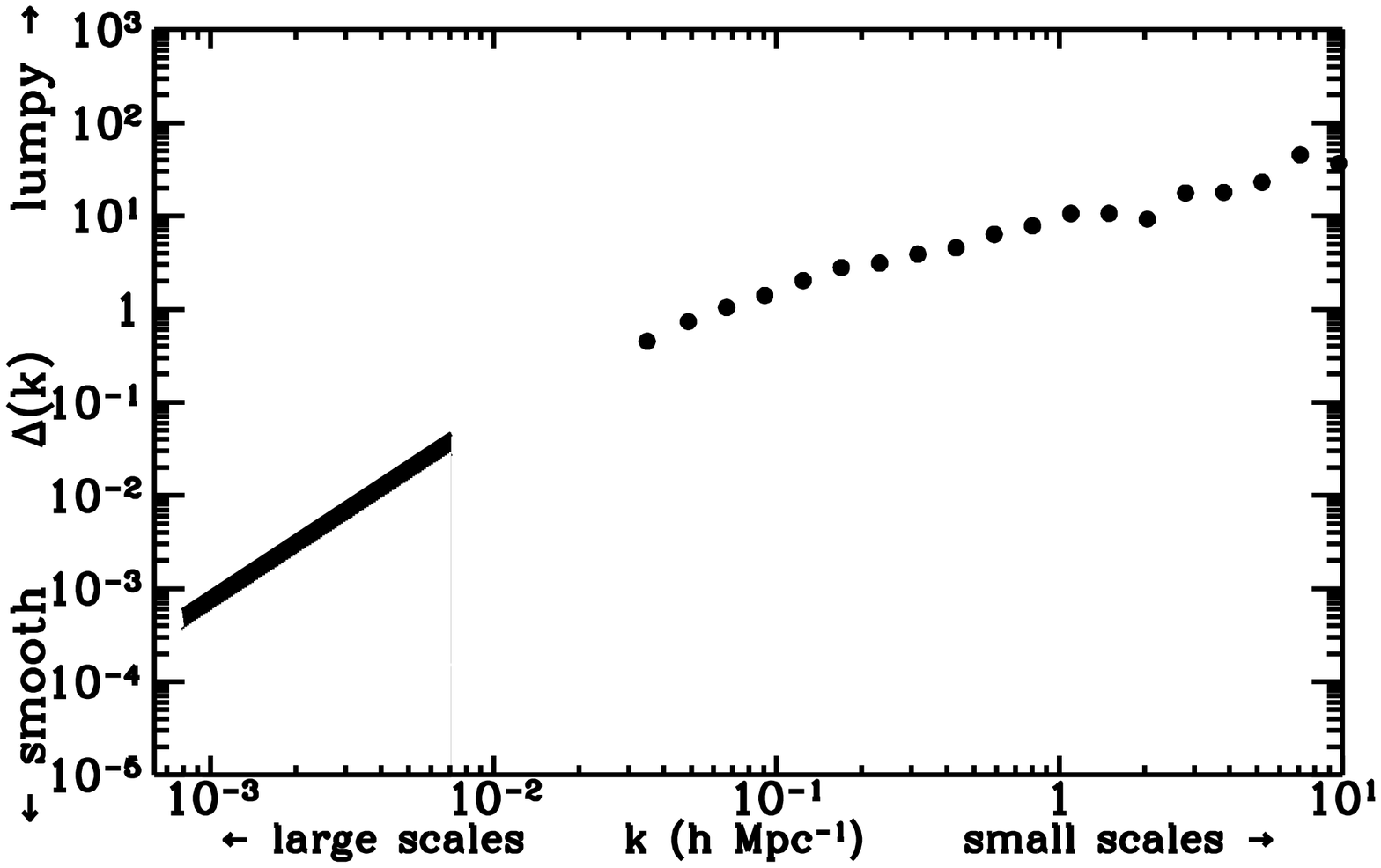}
\caption[fig4]{\label{p_k_gups} The ``grand unified'' power spectrum,
including determinations from large-scale structure surveys (the
points), and deduced from CBR temperature fluctuations (the box).}
\end{figure}

The region of wavenumber and amplitude of the power spectrum probed by
COBE is illustrated in Fig.\ \ref{p_k_cbr}.  There are now
measurements of CBR fluctuations on smaller angular scale,
corresponding to larger $k$.

Finally, Fig.\ \ref{p_k_gups} combines information from both
large-scale structure surveys and CBR temperature fluctuations.  The
trend is obvious: on small distance scales the power spectrum is
``large,'' which implies a lot of structure.  Matter is clustered on
small scales.  But on ``large'' scales the power spectrum decreases.
As one examines the universe on larger scales, homogeneity and
isotropy becomes a better and better approximation.

The data shown is only illustrative of many data sets.  Although
combining different data sets is uncertain and risky (problems with
normalization, {\it etc.}) the qualitative features are the same.  Figure
\ref{p_k_gups} is best regarded as an impressionist representation of
the situation.

Another thing to keep in mind is that the power spectrum may not be
the entire story.  The power spectrum contains all statistical
information about the perturbations only if the fluctuations are
Gaussian.  This should be cause for concern, because even if the
initial perturbations are Gaussian, eventually they will become
non-Gaussian once the perturbations become nonlinear.  Also, the power
spectrum is not a useful discriminant for prominent features such as
walls, voids, filaments, {\it etc.}  In spite of its drawbacks, the power
spectrum is remarkably useful---if we can't get the power spectrum
right, then we are not on the right track.

Now we turn to an early-universe theory that can account for the power
spectrum: inflation

\section{Inflation}
One of the striking features of the CBR fluctuations is that they {\it
appear} to be noncausal.  The CBR fluctuations were largely imprinted
at the time of last-scattering, about 300,000 years after the bang.
However, there are fluctuations on length scales much larger than
3000,000 light years!  How could a causal process imprint fluctuations
on scales larger than the light-travel distance since the time of the
bang?  The answer is inflation, but to see how that works, let's
define the problem more exactly.

First consider the evolution of the Hubble radius with the scale
factor $a(t)$:\footnote{Here and throughout the paper ``RD'' is short
for radiation dominated, and ``MD'' implies matter dominated.}
\begin{equation}
R_H\equiv H^{-1} = \left(\frac{\dot{a}}{a}\right)^{-1} 
\propto \rho^{-1/2}\propto \left\{\begin{array}{ll} a^2 & \mbox {(RD)}\\ 
	a^{3/2} & \mbox {(MD)}. \end{array} \right. 
\end{equation}
In a $k=0$ matter-dominated universe the age is related to $H$ by $t =
(2/3)H^{-1}$, so $R_H=(3/2)t$.  In the early radiation-dominated
universe $t=(1/2)H^{-1}$, so $R_H=2t$.

On length scales smaller than $R_H$ it is possible to move material
around and make an imprint upon the universe.  Scales larger than
$R_H$ are ``beyond the Hubble radius,'' and the expansion of the
universe prevents the establishment of any perturbation on scales
larger than $R_H$.

Next consider the evolution of some physical length scale $\lambda$.
Clearly, any physical length scale changes in expansion in proportion
to $a(t)$.

Now let us form the dimensionless ratio $L\equiv \lambda/R_H$.  If $L$
is smaller than unity, the length scale is smaller than the Hubble radius and
it is possible to imagine some microphysical process establishing
perturbations on that scale, while if $L$ is larger than unity, no
microphysical process can account for perturbations on that scale.

\begin{figure}[t]
\centering
\leavevmode\epsfxsize=177pt  \epsfbox{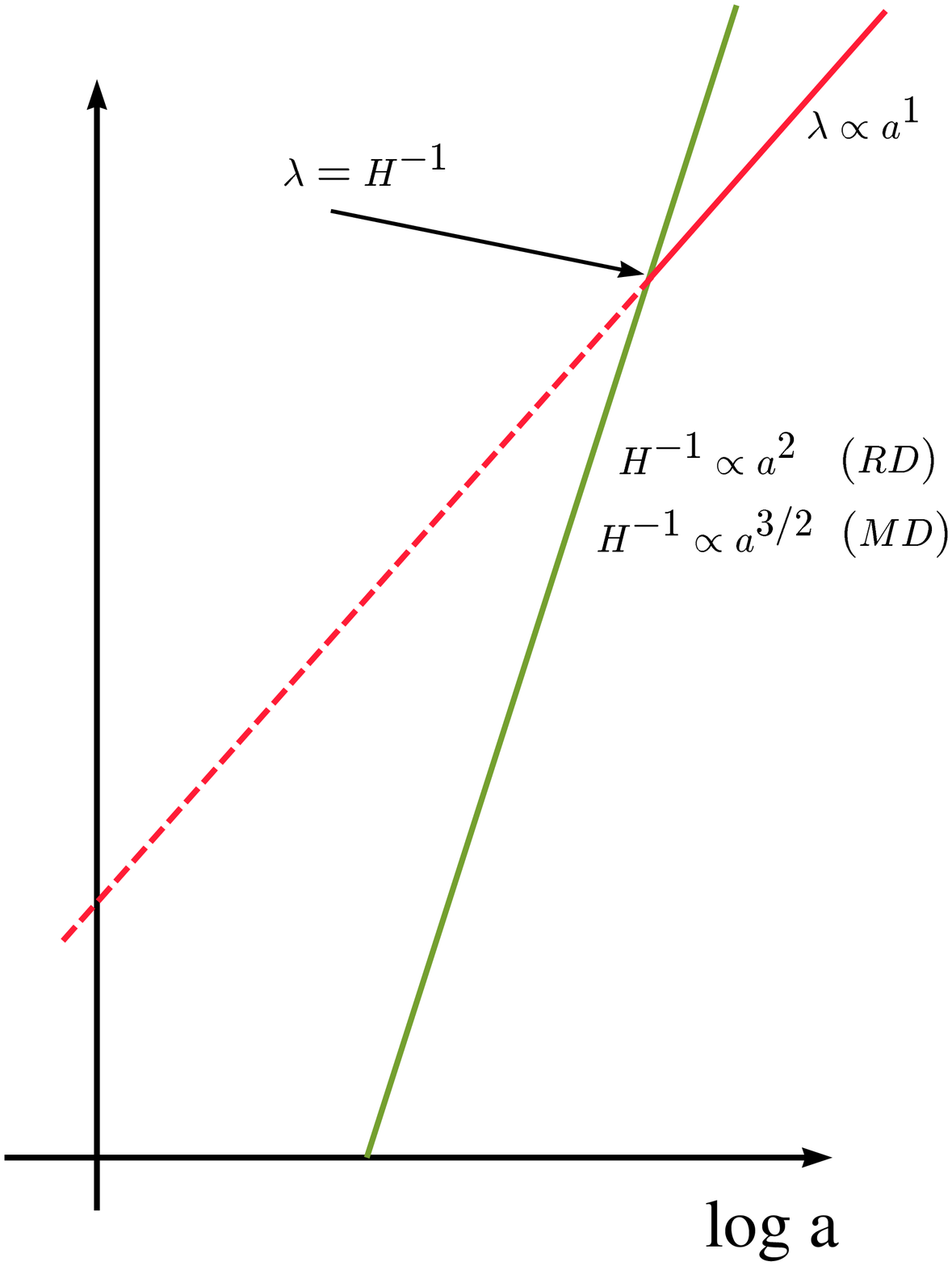}
\leavevmode\epsfxsize=177pt  \epsfbox{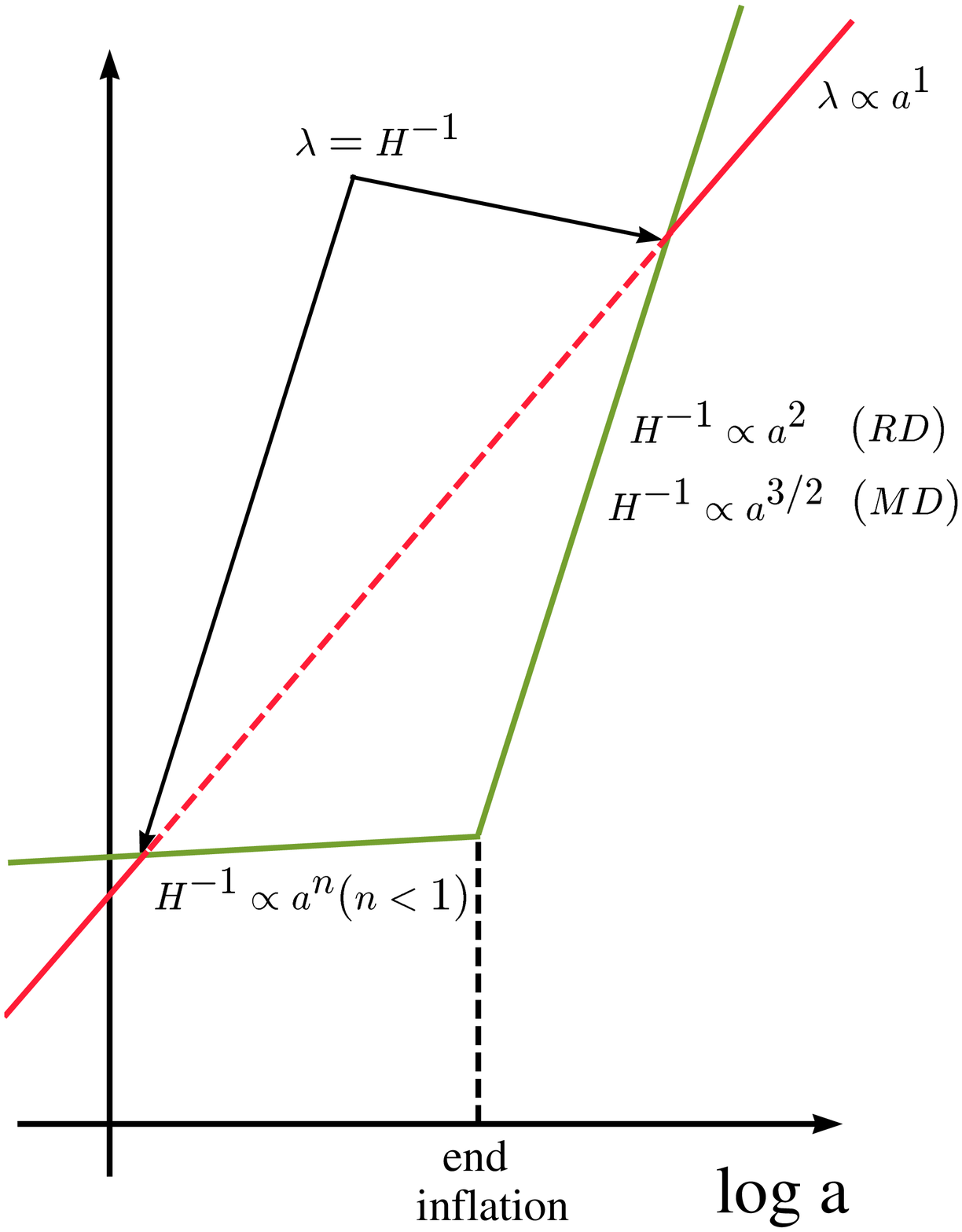}
\caption[fig5]{\label{a_hubble} Physical sizes increase as $a(t)$ in
the expanding universe.  The Hubble radius evolves as
$R_H=H^{-1}=(8\pi G \rho(a) /3)^{1/2}$.  In a radiation-dominated or
matter-dominated universe (left) any physical length scale $\lambda$
starts larger than $R_H$, then crosses the Hubble radius
($\lambda=H^{-1}$) only once. However, if there was a period of early
inflation (right) when $R_H$ increased more slowly than $a$, it is
possible for a physical length scale to start smaller than $R_H$,
become larger than $R_H$, and after inflation ends become once again
smaller than $R_H$.  Periods during which the scale is larger than the
Hubble radius are indicated by the dotted line.}
\end{figure}

Since $R_H=a/\dot{a}$, and $\lambda \propto a$, the ratio $L$ is
proportional to $\dot{a}$, and $\dot{L}$ scales as $\ddot{a}$, which
in turn is proportional to $-(\rho+3p)$.  There are two possible
scenarios for $\dot{L}$ depending upon the sign of $\rho+3p$:
\begin{equation}
\dot{L} \left\{  
\begin{array}{lll} <0 \rightarrow &  R_H \mbox{ grows faster than } \lambda, 
	& \mbox{happens for }\rho+3p>0 \\
<0 \rightarrow &  R_H \mbox{ grows more slowly than } \lambda, 
	& \mbox{happens for }\rho+3p  <0.
\end{array}
\right. 
\end{equation}
In the standard scenario, $\rho+3p>0$, $R_H$ grows faster than
$\lambda$.  This is illustrated by the left-hand side of
Fig. \ref{a_hubble}.

For illustration, let us take $\lambda$ to be the present length
$\lambda_8=8h^{-1}$Mpc, the scale beyond which perturbations today are
in the linear regime.  The physical length scale, which today is
$\lambda_8$, was smaller in the early universe by a factor of
$a(t)/a_0$, where $a_0$ is the scale factor today.  Today, the Hubble
radius is $H_0^{-1}\sim 3000 h^{-1}$Mpc.  Of course, today $\lambda_8$
is well within the current Hubble radius. But in the standard picture,
$R_H$ grows faster than $\lambda$, and there must therefore have been a time when
the comoving length scale that corresponds to $\lambda_8$ was larger
than $R_H$

Sometime during the early evolution of the universe the expansion was
such that $\ddot{a}>0$, which as we have just seen, requires an
unusual equation of state with $\rho+3p<0$.  This is referred to as
``accelerated expansion'' or ``inflation.''

\begin{table}[htb]
\begin{center}
\caption{\label{movimenti} Different epochs in the history of the
universe and the associated tempos of the expansion rate.}
\begin{tabular}{lllccc}
\hline
tempo & passage  & age & $\rho$ & $p$ & $\rho+3p$ \\
\hline
prestissimo & string dominated        &$<10^{-43}$s       & ?                                  & ?                & ? \\
presto         &  vacuum dominated (inflation)    &$\sim10^{-38}$s   & $\rho_V$                    & $-\rho_V$ & $-$\\
allegro        & matter dominated   &$\sim10^{-36}$s& $\rho_\phi$               & 0                &$+$ \\
andante     & radiation dominated      &$<10^{4}$yr        & $T^4$                        & $T^4/3$   &$+$\\
largo           & matter dominated       &$>10^{4}$yr    & $\rho_{\rm matter}$ & 0               &$+$\\
\hline
\end{tabular}
\end{center}
\end{table}

Including the inflationary phase, our best guess for the different
epochs in the history of the universe is given in Table
\ref{movimenti}.  There is basically nothing known about the stringy
phase, if indeed there was one.  The earliest phase we have
information about is the inflationary phase.  As we shall see, the
information we have is from the quantum fluctuations during inflation,
which were imprinted upon the metric, and can be observed as CBR
fluctuations and the departures from homogeneity and isotropy in the
matter distribution, {\it e.g.,} the power spectrum.  A lot of effort has
gone into studying the end of inflation.  It was likely that there was
a brief period of matter domination before the universe became
radiation dominated.  Very little is known about this period after
inflation.  Noteworthy events that might have occurred during this
phase include baryogenesis, phase transitions, and generation of dark
matter.  We do know that the universe was radiation dominated for
almost all of the first 10,000 years.  The best evidence of the
radiation-dominated era is primordial nucleosynthesis, which is a
relic of the radiation-dominated universe in the period 1 second to 3
minutes.  The earliest picture of the matter-dominated era is the CBR.

\begin{figure}[t]
\centering
\leavevmode\epsfxsize=177pt  \epsfbox{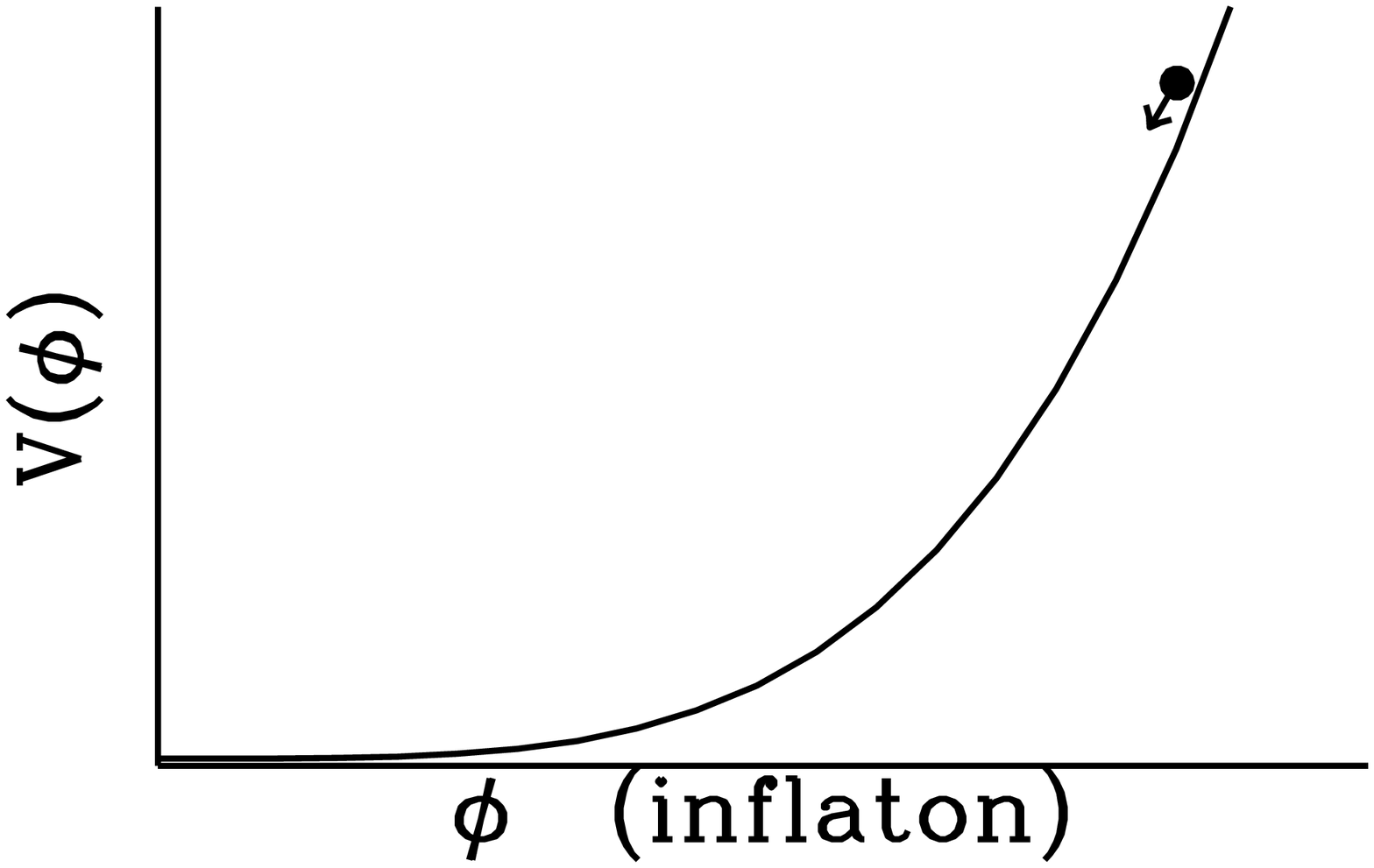}
\leavevmode\epsfxsize=177pt  \epsfbox{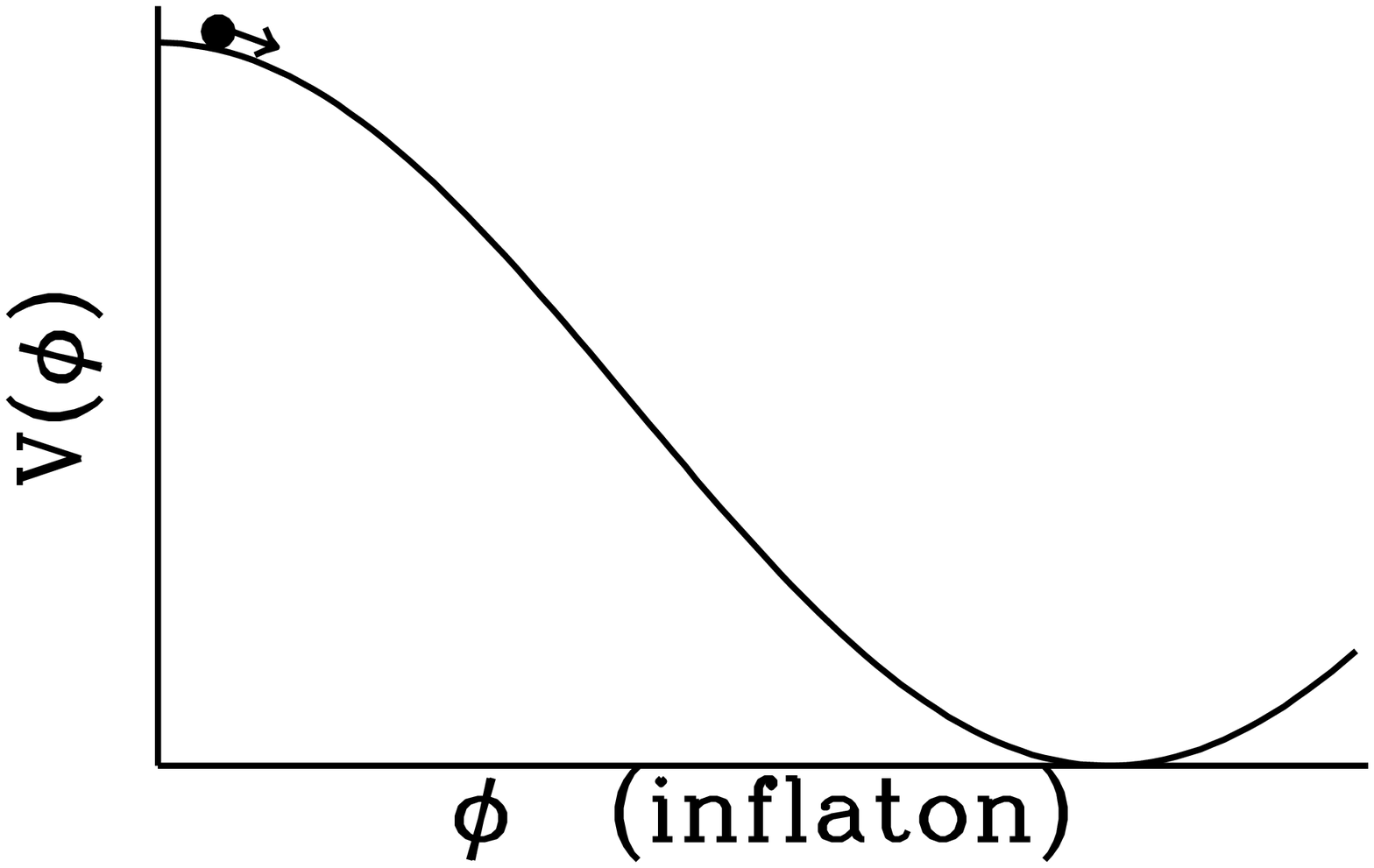}
\caption[fig6]{\label{large_small} Schematic illustrations of the
inflaton potential energy.  The potential on the left is a
``large-field'' model, where the inflaton field starts large and
evolves toward its minimum.  The right figure illustrates a
``small-field'' model.  A more accurate description of large-field and
small-field potential is the sign of the second derivative of the
potential: large-field models have $V''>0$ while small-field models
have $V''<0$.}
\end{figure}

Here, I am interested in events during the inflationary era.  The
first issue is how to imagine a universe dominated by vacuum energy
making a transition to a matter-dominated or radiation-dominated
universe.  A simple way to picture this is by the action of a scalar
field $\phi$ with potential $V(\phi)$.  Let's imagine the scalar field
is displaced from the minimum of its potential as illustrated in Fig.\
\ref{large_small}.  If the energy density of the universe is dominated
by the potential energy of the scalar field $\phi$, known as the {\it
inflaton}, then $\rho+3p$ will be negative.  The vacuum energy
disappears when the scalar field evolves to its minimum.

If the inflaton is completely decoupled, then it will oscillate about
the minimum of the potential, with the cycle-average of the energy
density decreasing as $a^{-3}$, {\it i.e.,} a matter-dominated universe.
But at the end of inflation the universe is cold and frozen in a
low-entropy state: the only degree of freedom is the zero-momentum
mode of the inflaton field.  It is necessary to ``defrost'' the
universe and turn it into a ``hot'' high-entropy universe with many
degrees of freedom in the radiation.  Exactly how this is accomplished
is still unclear.  It probably requires the inflaton field to be
coupled to other degrees of freedom, and as it oscillates, its energy
is converted to radiation either through incoherent decay, or through
a coherent process involving very complicated dynamics of coupled
oscillators with time-varying masses.  In either case, it is necessary
to extract the energy from the inflaton and convert it into radiation.

\subsection{QUANTUM FLUCTUATIONS}
During inflation there are quantum fluctuations in the inflaton field.
Since the total energy density of the universe is dominated by the
inflaton potential energy density, fluctuations in the inflaton field
lead to fluctuations in the energy density.  Because of the rapid
expansion of the universe during inflation, these fluctuations in the
energy density are frozen into super-Hubble-radius-size perturbations.
Later, in the radiation or matter-dominated era they will come within
the Hubble radius as if they were {\it noncausal} perturbations.

The spectrum and amplitude of perturbations depend upon the nature of
the inflaton potential.  Mukhanov \cite{mfb} has developed a very nice
formalism for the calculation of density perturbations.  One starts
with the action for gravity (the Einstein--Hilbert action) plus a
minimally-coupled scalar inflaton field $\phi$:
\begin{equation}
S = -\int d^4\!x \ \sqrt{-g} \left[ \frac{m^2_{Pl}}{16\pi}R - 
\frac{1}{2}g^{\mu\nu}\partial_\mu\phi\partial_\nu\phi  + V(\phi) \right] \  .
\end{equation} 
Here $R$ is the Ricci curvature scalar.  Quantum fluctuations result
in perturbations in the metric tensor
\begin{eqnarray}
g_{\mu\nu} & \rightarrow & g_{\mu\nu}^{FRW} + \delta  g_{\mu\nu} \nonumber \\
\phi & \rightarrow & \phi_0 + \delta \phi \ , 
\end{eqnarray}
where $ g_{\mu\nu}^{FRW}$ is the Friedmann--Robertson--Walker metric,
and $\phi_0(t) $ is the classical solution for the homogeneous,
isotropic evolution of the inflaton.  The action describing the
dynamics of the small perturbations can be written as
\begin{equation}
\delta_2S = \frac{1}{2}\int d^4\!x \ \left[ \partial_\mu u \partial^\mu u 
+ z^{-1}\frac{d^2z}{d\tau^2}\ u^2    \right]\ ;  \qquad z=a\dot{\phi}/H \  ,
\end{equation} 
{\it i.e.,} the action in conformal time $\tau$ ($d\tau^2 = a^2(t)dt^2)$ for
a scalar field in Minkowski space, with mass-squared $
m_u^2=-z^{-1}d^2z/d\tau^2$.  Here, the scalar field $u$ is a
combination of metric fluctuations $\delta g_{\mu\nu}$ and scalar
field fluctuations $\delta \phi$.  This scalar field is related to the
amplitude of the density perturbation.

The simple matter of calculating the perturbation spectrum for a
noninteracting scalar field in Minkowski space will give the amplitude
and spectrum of the density perturbations.  The problem is that the
solution to the field equations depends upon the background field
evolution through the dependence of the mass of the field upon $z$.
Different choices for the inflaton potential $V(\phi)$ results in
different background field evolutions, and hence, different spectra
and amplitudes for the density perturbations.

Before proceeding, now is a useful time to remark that in addition to
scalar density perturbations, there are also fluctuations in the
transverse, traceless component of the spatial part of the metric.
These fluctuations (known as tensor fluctuations) can be thought of as
a background of gravitons.

Although the scalar and tensor spectra depend upon $V(\phi)$, for most
potentials they can be characterized by $Q_{RMS}^{PS}$ (the amplitude
of the scalar and tensor spectra on large length scales added in
quadrature), $n$ (the scalar spectral index describing the best
power-law fit of the primordial scalar spectrum), $r$ (the ratio of
the tensor-to-scalar contribution to $C_2$ in the angular power
spectrum), and $n_T$ ( the tensor spectral index describing the best
power-law fit of the primordial tensor spectrum).  For single-field,
slow-roll inflation models, there is a relationship between $n_T$ and
$r$, so in fact there are only three independent variables.
Furthermore, the amplitude of the fluctuations often depends upon a
free parameter in the potential, and the spectra are normalized by
$Q_{RMS}^{PS}$.  This leads to a characterization of a wide-range of
inflaton potentials in terms of two numbers, $n$ and $r$.

In addition to the primordial spectrum characterized by $n$ and $r$,
in order to compare to data it is necessary to specify cosmological
parameters ($H_0$, the present expansion rate; $\Omega_0$, the ratio
of the present mass-energy density to the critical density---a
spatially flat universe has $\Omega_0=1$; $\Omega_B$, the ratio of the
present baryon density to the critical density; $\Omega_{DM}$ the
ratio of the present dark-matter density to the critical density; and
$\Lambda$, the value of the cosmological constant), as well as the
nature of the dark matter.

 The specification of the dark matter is by how ``hot'' the dark
 matter was when the universe first became matter dominated.  If the
 dark matter was really slow at that time, then it is referred to as
 cold dark matter.  If the dark matter was reasonably hot when the
 universe became matter dominated, then it is called hot dark matter.
 Finally, the intermediate case is called warm dark matter.  Neutrinos
 with a mass in the range 1 eV to a few dozen eV would be hot dark
 matter.  Light gravitinos, as appear in gauge-mediated supersymmetry
 breaking schemes, is an example of warm dark matter.  By far the most
 popular dark matter candidate is cold dark matter.  Examples of cold
 dark matter are neutralinos and axions.

\section{The Flavor of the Month}
There are exactly 31 different combinations of $n$, $r$, cosmological
parameters, and dark matter mixes.\footnote{This statement clearly is not
true.}  For this reason, different cosmological models are like
flavors of ice cream at Baskin Robbins.  There is always a flavor of
the month that everyone seems to like.  Flavors come in and out of
taste/fashion, with some adherents always choosing the same, while
others like to sample a wide variety.  A menu of the six most popular
flavors are given in Table \ref{baskin_robbins}.

\begin{table}[htb]
\begin{center}
\caption{\label{baskin_robbins} Different flavors of cosmological models.}
\begin{tabular}{lclcccccc}
\hline
flavor                     & $n$   & $r$ &                 $H_0$                   & $\Omega_0$ & $\Omega_B$ &$\Omega_{COLD}$ & $\Omega_{HOT}$ & $\Omega_\Lambda$ \\
\hline
CDM                     & 1       & 0   & 50 km s$^{-1}$Mpc$^{-1}$ & 1    & 0.05 & 0.95 &   0 & 0       \\
HDM                     & 1       & 0   & 50 km s$^{-1}$Mpc$^{-1}$ & 1    & 0.05 & 0    & 0.95 & 0        \\
MDM                     & 1       & 0   & 50 km s$^{-1}$Mpc$^{-1}$ & 1    & 0.10 &  0.70 & 0.20 & 0   \\
TCDM                    & 0.8  & 0   & 50 km s$^{-1}$Mpc$^{-1}$ & 1    & 0.05 & 0.95&  0 & 0   \\
OCDM                  & 1       & 0   & 50 km s$^{-1}$Mpc$^{-1}$ & 0.5 & 0.05 & 0.45&  0 & 0  \\
$\Lambda$CDM  & 1       & 0   & 50 km s$^{-1}$Mpc$^{-1}$ & 1    & 0.05 &  0.45  &  0 & 0.50  \\
\hline
\end{tabular}
\end{center}
\end{table}

\begin{figure}[t]
\centering
\leavevmode\epsfxsize=350pt  \epsfbox{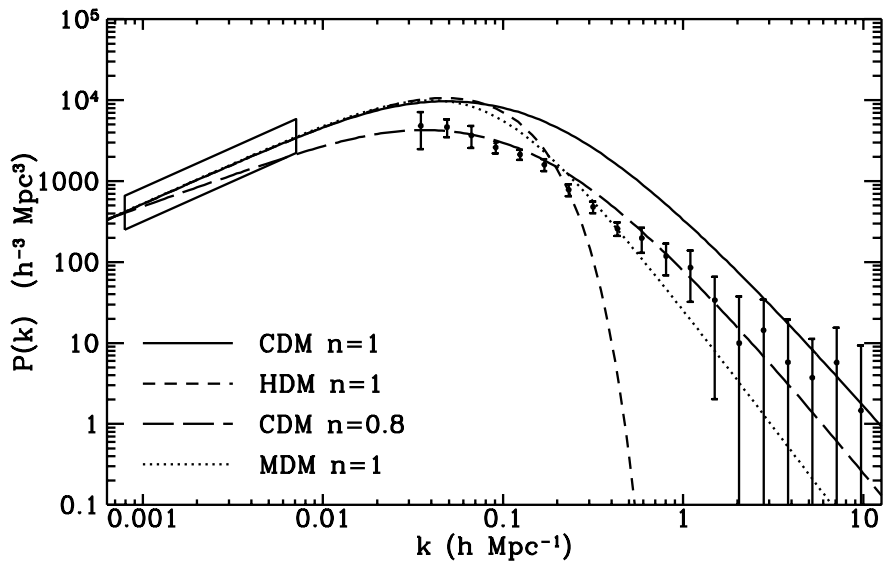}
\caption[fig5]{\label{comparison} The empirically determined power
spectrum of density perturbations and the (linear-theory) predictions
of several models.   The models shown are cold dark matter; hot dark
matter; tilted cold dark matter; and mixed dark matter.}
\end{figure}

Obviously, other combinations are possible.  A comparison of the
power spectrum in these models to our impressionist version of the
observationally determined power spectrum is shown in Fig.\
\ref{comparison}. Obviously CDM has too much power on small scales.
Hot dark matter is a disaster because it has {\it no} power on small
scales.  Tilted dark matter does better than CDM.  Mixed dark matter
does somewhat better, as does $\Lambda$ dark matter (not shown).
Rather than $\chi$-by-eye, I quote the results of one statistical
analysis, including many data sets, in Table \ref{gs}
\cite{erik_joe}.

\begin{table}[htb]
\begin{center}
\caption{\label{gs} One analysis of the comparison of data and models.}
\begin{tabular}{cc}
\hline
model                     & $\chi^2$/{\it d.o.f. }   \\
\hline
CDM                      & 3.8                                \\
TCDM                    & 2.1                                \\
$\Lambda$CDM  & 1.9                                  \\
OCDM                   & 1.8                                 \\
MDM                      & 1.2                                 \\
\hline
\end{tabular}
\end{center}
\end{table}

My reading of the comparison between data and experiment is that the
results of Table \ref{gs} should be regarded as a {\it relative}
measure of the agreement between models and present data.  For instance,
it is fair to say that MDM is a much better fit than CDM.  But one should be 
very careful before rejecting a model based upon these numbers.

Although one might get the best $\chi^2$ with a model having 10\%
baryons, 30\% cold dark matter, 30\% hot dark matter with 15\% each in
two species of neutrinos, 20\% cosmological constant, 10\% warm dark
mater, and seasoned with a little tilt, it doesn't mean that is the
way the universe is constructed.

Clearly, what is needed are better observations: finer-scale
observations of CBR fluctuations, as well as larger-scale
determinations of the power spectrum from large-scale structure
surveys.  In the next few years such experiments will be done.

There is now an aggressive campaign to measure CBR anisotropies on
fine angular scales.  The culmination of this program will be the
launch of two satellites---MAP by NASA and Planck by ESA.

Large-scale structure surveys are also progressing.  The largest
(three-dimensional) survey to date is the Las Campanas Redshift
Survey, containing the three-dimensional location of over 30,000
galaxies.  In the next two years another survey, called 2dF, will be
completed.  This survey will have about 100,000 galaxies in its
catalog.  Finally, the Sloan Digital Sky Survey will map $\pi$
steradians of the north galactic cap and find the location of
1,000,000 galaxies, along with 150,000 quasars.

By the time these experiments/observations are complete we will be in
the age of precision cosmology, and we should really be able to
compare theory and observation.

The remaining missing piece of the puzzle may be the identity of the
dark matter.

\section{Dark Matter}

\subsection{wimpy thermal relics}

In this school, the matter of neutrino masses has been reviewed in
great detail.  A neutrino of mass $m_\nu$ contributes to $\Omega_\nu h^2$ 
an amount 
\begin{equation}
\Omega_\nu h^2 = \left( \frac{m_\nu}{92{\rm eV}} \right) \ .
\end{equation}
If the mass of the neutrino is significantly less than 0.1 eV, then
its contribution to $\Omega_0$ is dynamically unimportant.

More promising than neutrino hot dark matter is cold dark matter.  The
most promising candidate for cold dark matter is the lightest
supersymmetric particle, presumably a neutralino.  Neutralino dark
matter has been well studied and reviewed \cite{vel}.

The next most popular dark-matter candidate is the axion.  Although the
axion is very light, since its origin is from a condensate, it is very
cold.  Axion dark matter has also been well studied and well reviewed
\cite{axionreview}.

There are presently several experiments searching for cosmic
neutralinos and cosmic axions.  Both types of searches seem
sensitive enough to discover the relic dark matter, although it will
take quite some time (and probably another generation of experiments)
to completely cover the parameter space.

Neutralinos are an example of a thermal relic.  A thermal relic is
assumed to be in local thermodynamic equilibrium (LTE) at early times.
The {\it equilibrium} abundance of a particle, say relative to the
entropy density, depends upon the ratio of the mass of the particle to
the temperature.  If we define the variables $Y\equiv n_X/s$ and
$x=M_X/T$, where $n_X$ is the number density of WIMP (weakly interacting 
massive particle) $X$ with mass
$M_X$ and $s \sim T^3$ is the entropy density, $Y\propto \exp(-x)$ for
$x\gg 1$, while $Y\sim$ constant for $x\ll 1$.  

\begin{figure}[t]
\centering
\leavevmode\epsfxsize=350pt  \epsfbox{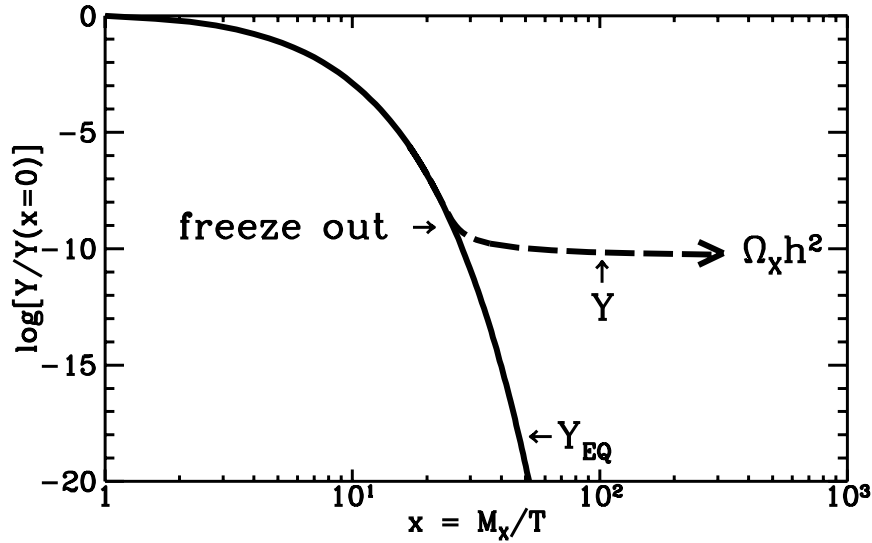}
\caption[fig8]{\label{thermal} A thermal relic starts in LTE at $T\gg M_X$.
When the rates keeping the relic in chemical equilibrium become smaller
than the expansion rate, the density of the relic relative to the entropy
density becomes constant.  This is known as {\it freeze out.}}
\end{figure}

A particle will track its equilibrium abundance so long as reactions
which keep the particle in chemical equilibrium can proceed rapidly
enough.  Here, rapidly enough means on a timescale more rapid than the
expansion rate of the universe $H$.  When the reactions becomes
slower than the expansion rate, then the particle can no longer track
its equilibrium value and thereafter $Y$ is constant.  When this
occurs, the particle is said to be ``frozen out.''  A schematic
illustration of this is given in Fig.\ \ref{thermal}.

The more strongly interacting the particle, the longer it stays in
LTE, and the smaller its freeze-out abundance.  Thus, the more weakly
interacting the particle, the larger its present abundance. The
freeze-out value of $Y$ is related to the mass of the particle and its
annihilation cross section (here characterized by $\sigma_0$) by
\begin{equation}
Y \propto \frac{1}{M_X m_{Pl} \sigma_0} \ .
\end{equation}
Since the contribution to $\Omega$ is proportional to $M_Xn_X$, which
in turn is proportional to $M_XY$, the present contribution to $\Omega$
from a thermal relic is (to first approximation) {\em independent} of
the mass, and only depends on the mass
indirectly through the dependence of the annihilation cross section on
mass. The largest that the annihilation cross section can be is roughly
$M_X^{-2}$.  This implies that large-mass WIMPS would have such a
small annihilation cross section that their present abundance would be
too large.  Thus, one expects a maximum mass for a thermal WIMP, which
turns out to be a few hundred TeV.

The mass of WIMPS usually considered for dark matter run from a
microvolt for axions to several dozen GeV for neutralinos.  With the
exception of massive magnetic monopoles, the possibility
of dark matter particles of GUT-scale mass is not usually considered,
because thermal relics of this mass would be expected to be
over abundant by several orders of magnitude.  

Recently, the idea that dark matter may be {\it supermassive} has
received a lot of attention.  Since wimpy little dark matter particles
with mass less than a TeV are called WIMPS, dark matter particles of
really hefty mass of $10^{12}$ to $10^{16}$ GeV seem to be more than
WIMPS, so they are referred to as WIMPZILLAS.

\subsection{WIMPZILLAS---SIZE DOES MATTER}
The simple assumption that the dark matter (DM) is a thermal relic is surprisingly
restrictive.  The limit $\Omega_X \simlt 1$ implies that the mass of a
DM relic must be less than about 500 TeV \cite{griestkam}.  The standard
lore is that the hunt for DM should concentrate on particles with mass
of the order of the weak scale and with interactions with ordinary
matter on the scale of the weak force. This has been the driving force
behind the vast effort in DM detectors.

But recent developments in understanding how
matter is created in the early universe suggests the possibility
that DM  might be naturally composed of {\it
nonthermal} supermassive states.  The supermassive dark matter (WIMPZILLA)
$X$ may have a mass many orders of magnitude larger than the
weak scale, possibly as large as the Grand Unified Theory (GUT)
scale.  It is very intriguing that these
considerations resurrect the possibility that the dark matter might be
charged or even strongly interacting!

The second condition for WIMPZILLAS is that the particle must not have been
in equilibrium when it froze out ({\it i.e.,} it is not a thermal
relic), otherwise $\Omega_X$ would be larger than one.
 A sufficient condition for nonequilibrium is that the
annihilation rate (per particle) must be smaller than the expansion
rate: $n\sigma|v|<H$, where $n$ is the number density, $\sigma |v|$ is
the annihilation rate times the M{\o}ller flux factor, and $H$ is the
expansion rate.  Conversely, if the WIMPZILLA was created at some
temperature $T_*$ {\it and} $\Omega_X<1$, then it is easy to show that
it could not have attained equilibrium.  To see this, assume $X$s
were created in a radiation-dominated universe at temperature $T_*$.
Then $\Omega_X$ is given by $\Omega_X =
\Omega_\gamma(T_*/T_0)M_Xn_X(T_*)/\rho_\gamma(T_*)$, where $T_0$ is
the present temperature (ignoring dimensionless
factors of order unity.)  Using the fact that $\rho_\gamma(T_*) =
H(T_*) M_{Pl} T_*^2$, we find $n_X(T_*)/H(T_*) =
(\Omega_X/\Omega_\gamma) T_0 M_{Pl} T_*/M_X$.  We may safely take the
limit $\sigma |v| < M_X^{-2}$, so $n_X(T_*)\sigma |v| / H(T_*)$ must
be less than $(\Omega_X / \Omega_\gamma) T_0 M_{Pl} T_* / M_X^3$.
Thus, the requirement for nonequilibrium is
\begin{equation}
\left( \frac{200\,{\rm TeV}}{M_X}\right)^2 \left( \frac{T_*}{M_X}
\right) < 1 \ .
\end{equation}
This implies that if a nonrelativistic particle with $M_X\simgt 200$
TeV was created at $T_*<M_X$ with a density low enough to result in
$\Omega_X \simlt 1$, then its abundance must have been so small that
it never attained equilibrium. Therefore, if there is some way to
create WIMPZILLAS in the correct abundance to give $\Omega_X\sim 1$,
nonequilibrium is guaranteed.

An attractive origin for WIMPZILLAS is during the defrosting phase after
inflation.  It is important to realize that it is not necessary to
convert a significant fraction of the available energy into massive
particles; in fact, it must be an infinitesimal amount.  If a fraction
$\epsilon$ of the available energy density is in the form of a
massive, stable $X$ particle, then $\Omega_X = \epsilon \Omega_\gamma
(T_{RH}/T_0)$, where $T_{RH}$ is the ``reheat'' temperature.  For
$\Omega_X=1$, this leads to the limit $\epsilon \simlt 10^{-17}
(10^9\, {\rm GeV}/T_{RH})$.  

In one extreme we might assume that the vacuum energy of inflation is
immediately converted to radiation, resulting in a reheat temperature
$T_{RH}$.  In this case $\Omega_X $ can be calculated by integrating
the Boltzmann equation with initial condition $N_X=0$ at $T=T_{RH}$.
One expects the $X$ density to be suppressed by $\exp(-2M_X/T_{RH})$;
indeed, one finds $\Omega_X \sim 1$ for $M_X/T_{RH} \sim 25 +
0.5\ln(M_X^2\langle \sigma |v|\rangle)$, in agreement with previous
estimates \cite{vadimvaleri} that for $T_{RH}\sim10^9$GeV, the WIMPZILLA
mass would be about $2.5\times10^{10}$GeV.

A second (and more plausible) scenario is that reheating is not
instantaneous, but is the result of the decay of the inflaton
field.  In this approach the radiation is produced as the inflaton 
decays.  The WIMPZILLA density is
found by solving the coupled system of equations for the inflaton
field energy, the radiation density, and the WIMPZILLA mass density.  The
calculation has been recently reported in Ref.\ \cite{CKRII}, with
result $\Omega_X \sim M_X^2\langle \sigma |v|\rangle
 (2000T_{RH}/M_X)^7$.  For a reheat temperature as
low as $10^9$GeV, a particle of mass $10^{13}$GeV can be produced in
sufficient abundance to give $\Omega_X \sim 1$.

\begin{figure}
\centerline{ \epsfxsize=350pt  \epsfbox{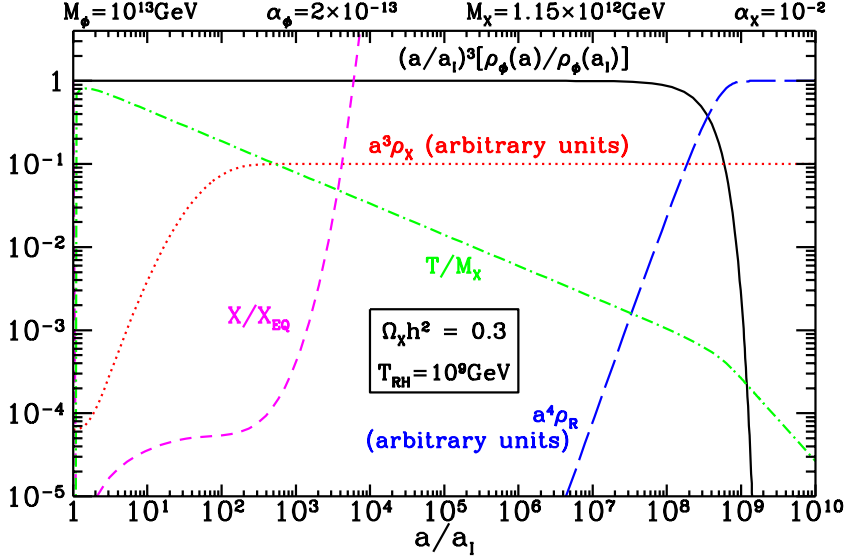} }
\caption{\label{reheat} The evolution of energy densities and $T/M_X$ as a function
of the scale factor.   Also shown is $X/X_{EQ}$.}
\end{figure}

The large difference in WIMPZILLA masses in the two reheating scenarios
arises because the peak temperature is much larger in the second
scenario, even with identical $T_{RH}$.  Because the temperature
decreases as $a^{-3/8}$ ($a$ is the scale factor) during most of the
reheating period in the second scenario, it must have once been much
greater than $T_{RH}$.  The evolution of the temperature is given in
Fig.\ \ref{reheat}.  
  If we assume the radiation spectrum did not depart
grossly from thermal, the effective temperature having once been
larger than $T_{RH}$ implies that the density of particles with enough
energy to create WIMPZILLAS was larger.  Denoting as $T_2$ the maximum
effective temperature for the second scenario, we find $T_2/T_{RH}
\sim (M_\phi/\Gamma_\phi)^{1/4} \gg 1$, where $\Gamma_\phi$ is the
effective decay rate of the inflaton.  See \cite{CKRII} for details.

Another way to produce WIMPZILLAS after inflation is in a preliminary stage
of reheating called ``preheating'' \cite{explosive}, where nonlinear
quantum effects may lead to an extremely effective dissipational
dynamics and explosive particle production. Particles can be created
in a broad parametric resonance with a fraction of the energy stored
in the form of coherent inflaton oscillations at the end of inflation
released after only a dozen oscillation periods.  A crucial
observation for our discussion is that particles with mass up to
$10^{15}$ GeV may be created during preheating \cite{klr,kt,gut}, and
that their distribution is nonthermal. If these particles are stable,
they may be good candidates for WIMPZILLAS.

To study how the creation of WIMPZILLAS takes place in preheating, let us
take the simplest chaotic inflation potential:
$V(\phi)=M_\phi^2\phi^2/2$ with $M_\phi\sim 10^{13}$ GeV.  We assume
that the interaction term between the WIMPZILLA and the inflaton field is of
the type $g^2\phi^2|X|^2$.  Quantum fluctuations of the $X$ field with
momentum $\vec{k}$ during preheating {\em approximately} obey the
Mathieu equation, $ X_k'' + [A(k) - 2q\cos2z]X_k =0$, where $q = g^2
\phi^2 / 4 M_\phi^2$, $A(k) = (k^2 + M_X^2) / M_\phi^2 + 2q$ (primes
denotes differentiation with respect to $z=M_\phi t$).  Particle
production occurs above the line $A = 2 q$ in an instability strip of
width scaling as $q^{1/2}$ for large $q$.  The condition for broad
resonance, $A-2q \simlt q^{1/2}$ \cite{klr}, becomes $(k^2 +
M^2_X)/M_\phi^2 \simlt g \bar\phi / M_\phi$, which yields $E_X^2 =
{k^2 + M^2_X } \simlt g \bar\phi M_\phi $ for the typical energy of
particles produced in preheating. Here $\bar\phi$ is the amplitude of
the oscillating inflaton field \cite{explosive}.  The resulting
estimate for the typical energy of particles at the end of the broad
resonance regime for $M_\phi \sim 10^{-6} M_{\rm Pl}$ is $E_X \sim
10^{-1} g^{1/2}\sqrt { M_\phi M_{\rm Pl}} \sim g^{1/2} 10^{15}$ GeV.
Supermassive $X$ bosons can be produced by the broad parametric
resonance for $E_X > M_X$, which leads to the estimate that $X$
production will be possible if $M_X < g^{1/2} 10^{15}$ GeV. For $g^2
\sim 1$ one would have copious production of $X$ particles as heavy as
$10^{15}$GeV, {\it i.e.}, 100 times greater than the inflaton mass,
which may be many orders of magnitude greater than the reheat
temperature. Scatterings of $X$ fluctuations off the zero mode of the
inflaton field considerably limits the maximum magnitude of $X$
fluctuations to be $\langle X^2\rangle_{\rm max} \approx M_\phi^2/g^2$
\cite{KT3}.  For example, $\langle X^2\rangle_{\rm max} \simlt
10^{-10} M_{\rm Pl}^2$ in the case $M_X = 10\:M_\phi$. This restricts
the corresponding number density of created $X$-particles.

For a reheating temperature of the order of 100 GeV, the present
abundance of WIMPZILLAS with mass $M_X\sim 10^{14}$ GeV is given by
$\Omega_{X}\sim 1$ if $\epsilon \sim 10^{-10}$. This small fraction
corresponds to $\langle X^2\rangle \sim 10^{-12} M_{\rm Pl}^2$ at the
end of the preheating stage, a value naturally achieved for WIMPZILLA mass
in the GUT range \cite{KT3}. The creation of WIMPZILLAS through preheating
and, therefore, the prediction of the present value of $\Omega_X$, is
very model dependent. The inflaton might preferably decay through
parametric resonance into very light boson fields so that the end of
the preheating stage and of the corresponding value of $\langle
X^2\rangle$ depends upon the coupling of the inflaton field not only
to the WIMPZILLA, but also to other degrees of freedom. It is
encouraging, however, that it is possible to produce supermassive
particles during preheating that are as massive as $10^{12}T_{RH}$.
Details of WIMPZILLA production in preheating can be found in
\cite{dansthesis}.

Another possibility which has been recently investigated is the
production of very massive particles by gravitational mechanisms
\cite{ckr,igor}. In particular, the desired abundance of WIMPZILLAS may be
generated during the transition from the inflationary phase to a
matter/radiation dominated phase as the result of the expansion of the
background spacetime acting on vacuum quantum fluctuations of the dark
matter field \cite{ckr}. A crucial side-effect of the inflationary
scenarios is the generation of density perturbations. A related
effect, which does not seem to have attracted much attention, is the
possibility of producing matter fields due to the rapid change in the
evolution of the scale factor around the end of inflation. Contrary to
the first effect, the second one contributes to the homogeneous
background energy density that drives the cosmic expansion, and is
essentially the familiar ``particle production'' effect of
relativistic field theory in external fields.

Very massive particles may be created in a nonthermal state with
sufficient abundance to achieve critical density today by the
classical gravitational effect on the vacuum state at the end of
inflation. Mechanically, the particle creation scenario is similar to
the inflationary generation of gravitational perturbations that seed
the formation of large-scale structures.  However, the quantum
generation of energy density fluctuations from inflation is associated
with the inflaton field, which dominated the mass density of the
universe, and not a generic sub-dominant scalar field.

If $0.04 \simlt M_{X}/H_e \simlt 2$ \cite{ckr}, where $H_e$ is the
Hubble constant at the end of inflation, DM produced gravitationally
can have a density today of the order of the critical density. This
result is quite robust with respect to the fine details of the
transition between the inflationary phase and the matter-dominated
phase.  The only requirement is that
\begin{equation}
\left(\frac{H_e}{10^{-6}M_{Pl}}\right)^2\left(\frac{T_{RH}}{10^9{\rm GeV}}\right)\simgt10^{-2}\ .
\end{equation}
The observation of anisotropy in the cosmic background radiation
does not fix $H_e$ uniquely, but using $T_{RH}\simlt\sqrt{M_{Pl}H_e}$,
we find that the  mechanism is effective only when $H_e \simgt 10^9$GeV 
(or, $M_X\simgt  10^8$GeV).

The distinguishing feature of this mechanism \cite{ckr} is
the capability of generating particles with mass of the order of the
inflaton mass even when the WIMPZILLA interacts only extremely weakly (or
not at all!) with other particles, including the inflaton.   This
feature makes the gravitational production mechanism quite model
independent and, therefore, more appealing to us than the one
occurring at preheating.

WIMPZILLAS can also be produced in theories where
inflation is completed by a first-order phase transition \cite{ls}. In
these scenarios, the universe decays from its false vacuum state by
bubble nucleation \cite{guth}.  When bubbles form, the energy of the
false vacuum is entirely transformed into potential energy in the
bubble walls, but as the bubbles expand, more and more of their energy
becomes kinetic and the walls become highly relativistic. Eventually
the bubble walls collide.

During collisions, the walls oscillate through each other \cite{moss}
and the kinetic energy is dispersed into low-energy scalar waves
\cite{moss,wat}.  If these soft scalar quanta carry quantum numbers
associated with some spontaneously broken symmetry, they may even lead
to the phenomenon of nonthermal symmetry restoration \cite{col}.  We
are, however, more interested in the fate of the potential energy of
the walls, $M_P = 4\pi\eta R^2$, where $\eta$ is the energy per unit
area of the bubble with radius $R$.  The bubble walls can be imagined
as a coherent state of inflaton particles, so that the typical energy
$E$ of the products of their decays is simply the inverse thickness of
the wall, $E\sim \Delta^{-1}$. If the bubble walls are highly
relativistic when they collide, there is the possibility of quantum
production of nonthermal particles with mass well above the mass
 of the inflaton field, up to energy $\Delta^{-1}=\gamma
M_\phi$, $\gamma$ being the relativistic Lorentz factor.

Suppose now that the WIMPZILLA is some fermionic degree of freedom $X$ and
that it couples to the inflaton field by the Yukawa coupling $g
\phi\overline{X}{X}$. One can treat $\phi$ (the bubbles or walls) as a
classical, external field and the WIMPZILLA as a quantum field in the
presence of this source. This amounts to ignoring the backreaction of
particle production on the evolution of the walls, but this is
certainly a good approximation in our case. The number of WIMPZILLA
particles created in the collisions from the wall's potential energy
is $N_X\sim f_X M_P/M_X$, where $f_X$ parametrizes the fraction of the
primary decay products that are WIMPZILLAS.  The
fraction $f_X$ will depend in general on the masses and the couplings
of a particular theory in question.  For the Yukawa coupling $g$, it
is $ f_X \simeq g^2 {\rm ln}\left(\gamma M_\phi/2 M_{X}\right)$
\cite{wat,mas}.  Supermassive particles in bubble collisions are
produced out of equilibrium and they never attain chemical
equilibrium. Assuming $T_{RH}\simeq 100$ GeV, the present abundance of
WIMPZILLAS is $\Omega_{X}\sim 1$ if $g\sim 10^{-5}\alpha^{1/2}$. Here
$\alpha^{-1}\ll 1$ denotes the fraction of the bubble energy at
nucleation which has remained in the form of potential energy at the
time of collision.  This simple analysis indicates that the correct
magnitude for the abundance of $X$ particles may be naturally obtained
in the process of reheating in theories where inflation is terminated
by bubble nucleation.

In conclusion,  a large fraction of the DM in the
universe may be made of WIMPZILLAS of mass
greatly exceed the electroweak scale---perhaps as large as the GUT
scale. This is made possible by the fact that the WIMPZILLAS were created in a
nonthermal state and never reached chemical equilibrium with the
primordial plasma.

\section*{Acknowledgements}  
This work was supported in part by the Department of Energy, as well
as NASA under grant number NAG5-7092.  The hospitality of Tom Ferbel
and the inquisitiveness of the students were greatly appreciated.

\end{document}